\documentclass[aps,prd,twocolumn,nofootinbib,superscriptaddress,preprintnumbers]{revtex4-2}

\usepackage{amsmath,amssymb,bm}
\usepackage{braket}
\usepackage{graphicx}
\usepackage{booktabs}
\usepackage{xcolor}
\usepackage[normalem]{ulem}
\usepackage[colorlinks=true,linkcolor=blue,citecolor=blue,urlcolor=blue]{hyperref}
\usepackage{orcidlink}

\DeclareMathOperator{\Tr}{Tr}
\newcommand{\dd}{\mathrm{d}}
\newcommand{\kin}{\mathrm{kin}}

\newcommand{\tot}{\mathrm{tot}}

\newcommand{\innerProd}[2]{\langle #1 | #2 \rangle}
\colorlet{BLUE}{blue}
\colorlet{RED}{red}

\DeclareRobustCommand{\ndel}[1]{\textcolor{gray}{\sout{#1}}}
\newcommand{\M}{\mathcal M}
\makeatletter
\newcommand{\setsectionprefixedsubsections}{%
  \renewcommand{\thesubsection}{\thesection.\Alph{subsection}}%
  \renewcommand{\p@subsection}{}%
}
\makeatother
\setsectionprefixedsubsections

\begin{document}

\title{Flavor--Kinetic Entanglement Production from Decay and Scattering at Finite Density}

\author{Zekun Li}
\email{lizekun@stu.pku.edu.cn}
\affiliation{School of Physics, Peking University, Beijing 100871, China}

\author{Jia Liu \orcidlink{0000-0001-7386-0253}}
\email{jialiu@pku.edu.cn}
\affiliation{School of Physics, Peking University, Beijing 100871, China}
\affiliation{State Key Laboratory of Nuclear Physics and Technology, Peking University, Beijing 100871, China}
\affiliation{Center for High Energy Physics, Peking University, Beijing 100871, China}

\author{Xiao-Ping Wang \orcidlink{0000-0002-2258-7741}}
\email{hcwangxiaoping@buaa.edu.cn}
\affiliation{School of Physics, Beihang University, Beijing 100083, China}

\author{Jing-Jun Zhang \orcidlink{0009-0007-5228-5959}}
\email{zhang\_jingjun@stu.pku.edu.cn}
\affiliation{School of Physics, Peking University, Beijing 100871, China}

\preprint{CPTNP-2026-023}

\begin{abstract}
We extend the scattering-entanglement dictionary to finite-density environments by investigating the flavor--kinetic bipartition of the Hilbert space. We show that tracing over kinematic degrees of freedom maps the total branch-changing transition probability directly onto the leading flavor--kinetic linear entanglement entropy.
At finite density, the vacuum branch-changing probability is replaced by an occupation-weighted collision probability, built from the same directed reaction-density kernel that enters the integrated Boltzmann equation. The resulting observable is the bath-averaged flavor--kinetic entanglement entropy of a pair sampled from the medium.
As a proof of principle, this framework is applied to an $O(N)$ singlet-scalar extended model to probe thermal phase transitions. In the examples studied, the resulting entanglement entropy serves as a collision-based phase-transition-type diagnostic, exhibiting a finite discontinuity across a first-order phase transition and a nonanalytic temperature derivative for continuous transitions. These examples suggest a novel way to characterize thermal phase structures, distinct from traditional thermodynamic order parameters.
\end{abstract}

\maketitle

\section{Introduction}
\label{sec:intro}

Scattering is the operational interface between quantum field theory and experiment. Conventionally, an $S$-matrix element is used to compute transition probabilities, cross sections, and reaction rates. The same $S$-matrix, however, is also a quantum map acting on a tensor-product Hilbert space, and therefore it can generate entanglement among particle labels, momenta, spins, helicities, flavors, and other internal quantum numbers. Since a scattering process maps a prepared in-state into a coherent superposition over resolved and unresolved final-state channels, the generated entanglement depends not only on the invariant amplitude but also on the Hilbert-space partition imposed by the measurement prescription. This viewpoint has led to a growing scattering-entanglement program in which entanglement generation is treated as a quantum-information observable associated with the same amplitudes that determine measurable transition probabilities~\cite{Seki:2014cgq,Peschanski:2016hgk,Carney:2016tcs,Cervera-Lierta:2017tdt,Fan:2017hcd,Fedida:2022izl,Low:2024mrk,Low:2024hvn,Aoude:2024xpx}.

This scattering perspective is part of a broader quantum-information program in particle physics. At colliders, unstable particles whose decay distributions retain spin-analyzing information provide experimentally reconstructible density matrices, making top-quark pairs, weak-boson pairs, Higgs decay products, and heavy-flavor systems natural laboratories for quantum tomography, entanglement witnesses, Bell inequalities, steering, discord, and new-physics-sensitive quantum observables~\cite{Barr:2024djo,Afik:2020onf,Fabbrichesi:2021npl,Severi:2021cnj,Afik:2022kwm,Afik:2022dgh,Aoude:2022imd,Aguilar-Saavedra:2022uye,Ashby-Pickering:2022umy,Cheng:2025zaw,Aguilar-Saavedra:2026wuq}. In particular, top-quark pairs have become a benchmark high-energy two-qubit system, with recent ATLAS and CMS measurements reporting quantum entanglement in $t\bar t$ production~\cite{ATLAS:2023fsd,CMS:2024pts}. Related developments include Bell and entanglement studies in Higgs and diboson processes, SMEFT tomography, $\tau$-pair and heavy-flavor decays, hyperon spin correlations, light-quark fragmentation, and small-$x$ hadronic structure~\cite{Barr:2021zcp,Aguilar-Saavedra:2022wam,Fabbrichesi:2023wvb,Aoude:2023hxv,Fabbrichesi:2023cev,Ehatht:2023zzt,Gabrielli:2024yci,Cheng:2025qit,Gong:2021bcp,Kharzeev:2017qzs}. These works emphasize that quantum-information observables are not merely formal diagnostics: in favorable channels they can be reconstructed from angular distributions, spin correlations, or final-state multiplicity information, and can therefore complement conventional cross-section and asymmetry measurements.

Complementary to this experimentally driven collider program, an amplitude-level line of work asks whether entanglement extrema or suppression encode parameters, mixing patterns, and symmetry structure.  Entanglement extrema have been used to characterize SM parameters and mixing data, including electroweak mixing~\cite{Cervera-Lierta:2017tdt}, leptonic CP violation~\cite{Quinta:2022sgq}, CKM/PMNS structures~\cite{Thaler:2024anb}, and Higgs-sector masses~\cite{Liu:2025iwh}.  Minimal or suppressed scattering entanglement has also been tied to emergent or enhanced symmetries in hadronic and scalar sectors, including Wigner-type spin--flavor symmetry, heavy-hadron symmetry patterns, Higgs alignment, mirror-sector structures, and related scalar-sector symmetry limits~\cite{Beane:2018oxh,Low:2021ufv,Liu:2022grf,Liu:2023bnr,Hu:2025jne,McGinnis:2025brt,McGinnis:2025xgt,Low:2026oyf,Sone:2026jmo,Low:2026evp,Carena:2023vjc,Carena:2025wyh,Kowalska:2024kbs,Kowalska:2025qmf}.  More broadly, scattering entanglement and related entropy measures have been studied in momentum-space, relativistic, bootstrap, perturbative, nonstabilizer, and flavored-scattering settings~\cite{Grignani:2016igg,Fan:2017mth,Bose:2020shm,Cheung:2023hkq,Sou:2025tyf,Lyu:2025lja,Gargalionis:2025iqs,Fabbrichesi:2026jvm,Cao:2026aye,Peschanski:2026edo,Gargalionis:2026onv}.

The relation between scattering entanglement and measurable scattering degrees of freedom (D.O.F.) has also been sharpened. Elastic-scattering entanglement can be expressed in terms of partial-wave data, elastic and total cross sections, and the finite-volume regularization of the two-body Hilbert space~\cite{Peschanski:2016hgk,Peschanski:2019yah}. In wave-packet formulations, the leading entanglement entropy for suitable $2\to2$ partitions is the corresponding elastic or semi-inclusive scattering probability, equivalently a cross section divided by the transverse wave-packet area~\cite{Low:2024mrk,Low:2024hvn}. This identifies the subsystem choice as an inclusive measurement prescription: after unobserved D.O.F. are traced out, the leading entanglement entropy is controlled by the summed transition probability selected by that bipartition.

Existing scattering-entanglement dictionaries are formulated primarily in vacuum. Realistic astrophysical and cosmological environments such as the early universe, however, are thermal and finite-density many-body systems. The appropriate microscopic language is finite-temperature and finite-density quantum field theory, where thermal propagators, real- or imaginary-time formalisms, distribution functions, and medium-dependent quasiparticle properties enter physical rates~\cite{Kapusta:2006pm,Laine:2016hma,Quiros:1999jp}. In kinetic descriptions, the microscopic amplitudes are further organized into collision operators or reaction densities: initial-state occupation numbers weight the sampled particles in the bath, final-state Bose enhancement or Pauli blocking modifies the available phase space, and medium corrections can enter through thermal masses, screening, damping, or effective couplings~\cite{Arnold:2002zm,Gondolo:1990dk}. Vacuum two-particle scattering therefore fails to capture the many-body factors that control macroscopic real-time evolution. Extending the scattering-entanglement dictionary to finite density is not just a theoretical curiosity, but a necessary step for applying quantum-information tools to cosmological, astrophysical, and non-equilibrium dynamics.

To discuss the quantum entanglement structure in a multi-particle process, we must carefully define the Hilbert-space partition. In this work, we focus on the flavor--kinetic bipartition $\mathcal{H}=\mathcal{H}_{\mathrm{f}}\otimes\mathcal{H}_{\rm kin}$, where a flavor branch denotes the discrete labels retained in the reduced subsystem, including species and, when relevant, color, spin, helicity, or flavor, while the kinetic subsystem consists of momentum configurations~\cite{Low:2024mrk,Low:2024hvn}. After tracing over the kinematic D.O.F., the mixedness of the reduced flavor density matrix defines the flavor--kinetic entanglement entropy $\mathcal{E}_{\rm fk}=1-\Tr_f\rho_f^2$. The central result of our finite-density extension is that $\mathcal{E}_{\rm fk}$ is directly governed by an occupation-weighted collision probability, schematically built from
\begin{equation}
    \mathcal{E}_{\rm fk}^{\rm FD} \propto \int d\Phi\, f_a f_b\, (1\pm f_c)(1\pm f_d)\, |\mathcal M_{\rm FD}|^2 .
    \label{eq:intro-fd-schematic}
\end{equation}
This is the same occupation-weighted directed reaction-density kernel that appears as a building block of the integrated Boltzmann equation. The event-level normalization is essential: the finite-density entanglement entropy is the bath average of event-level entanglement entropies for a single pair drawn from the medium, summed over unresolved kinematic configurations, whereas the Boltzmann equation further contracts the same kernel with gain-minus-loss and stoichiometric factors to evolve macroscopic number densities.

As an application, we use this finite-density entanglement entropy to probe thermal phase transitions in an $O(N)$ singlet-scalar model coupled to the Higgs sector. Singlet extensions of the scalar sector are among the minimal Higgs-portal frameworks capable of modifying the electroweak thermal history, for example, by strengthening the transition, generating tree-level barriers, or realizing multi-step thermal evolution~\cite{Espinosa:1993bs,Profumo:2007wc,Curtin:2014jma,Kurup:2017dzf,Niemi:2024axp}. Traditional phase-transition analyses typically rely on static thermodynamic effective potentials, order parameters, latent heat, or bubble-nucleation quantities. In contrast, the flavor--kinetic entanglement entropy proposed here is a collision-kernel-based dynamical observable. It monitors how microscopic finite-density scattering rates---and consequently entanglement generation between flavor and kinetic sectors---respond to the thermal background. In the examples studied below, the behavior of $\mathcal{E}_{\rm fk}$ near critical temperatures can diagnose the type of phase transition: first-order phase transitions (FOPTs) produce a finite jump in this entanglement entropy, while continuous transitions appear through a nonanalytic temperature derivative.

The remainder of this paper is organized as follows. In Sec.~\ref{sec:smatrix}, we define the flavor--kinetic partition and derive the leading-order decay and vacuum scattering relations from $S$-matrix unitarity. In Sec.~\ref{sec:finite-density}, we extend the construction to finite density and show that the leading flavor--kinetic entanglement entropy is governed by the directed reaction-density kernel that also enters the Boltzmann collision operator. In Sec.~\ref{sec:thermal-application}, we apply the formalism to an $O(N)$ scalar extended model and study the finite-density flavor--kinetic entanglement entropy across thermal phase transitions. Finally, conclusions and outlook are given in Sec.~\ref{sec:conclusions}.

\section{Flavor--Kinetic Linear Entanglement Entropy}
\label{sec:smatrix}

This section defines the flavor--kinetic entanglement entropy used throughout the paper and derives its leading contribution from the $S$-matrix. We first work in vacuum. Thus, for a fixed microscopic incoming branch, the outgoing state is pure in the full flavor--kinetic Hilbert space and mixedness appears only after the kinematic degrees of freedom are traced out.

\subsection{Hilbert-space partition and conventions}
\label{subsec:hilbert-space}

Previous studies of scattering entanglement have often focused on bipartitions that are natural for two-body final states, and the leading entanglement entropy for such partitions can be related to the corresponding scattering probability or cross section~\cite{Low:2024hvn,Low:2024mrk}. For a general final state with arbitrary multiplicity, however, a particle-by-particle bipartition is less canonical, and the full multipartite entanglement structure rapidly becomes partition-dependent and technically involved~\cite{Zeng:2015pxf}. We therefore use a bipartition that remains meaningful for arbitrary multiplicity: the discrete flavor branch is separated from the continuous kinematic configuration.

At fixed particle number $N$, we use the sectorwise decomposition
\begin{equation}
    \mathcal H^{(N)}
    \simeq
    \mathcal H_f^{(N)}
    \otimes
    \mathcal H_{\rm kin}^{(N)} .
\end{equation}
The full Fock space is then the direct sum over particle-number sectors,
\begin{equation}
    \mathcal H
    =
    \bigoplus_{N=0}^{\infty}
    \mathcal H^{(N)} .
\end{equation}
A flavor branch $\mathcal F$ denotes the discrete labels of the particles in the state, including species and any other discrete labels retained in the chosen description, such as spin, helicity, flavor, or color. The corresponding kinematic configuration is denoted by $p_{\mathcal F}=\{p_r\}_{r\in\mathcal F}$. We use the basis
\begin{equation}
    \ket{X_{\mathcal F};p_{\mathcal F}}
    \equiv
    \ket{X_{\mathcal F}}_f
    \otimes
    \ket{p_{\mathcal F}}_{\rm kin},
\end{equation}
or, equivalently,
\begin{equation}
    \ket{X_{\mathcal F};p_{\mathcal F}}
    =
    \ket{X_1;p_1}\cdots
    \ket{X_{N_{\mathcal F}};p_{N_{\mathcal F}}},
\end{equation}
with the Fock-space symmetrization or antisymmetrization understood. Identical-particle symmetry factors are included explicitly in the phase-space measures below.

For a one-particle state, we use the covariant normalization
\begin{equation}
    \innerProd{X_a;p_a}{X_b;p_b}
    =
    \delta_{ab}\,2E_a(2\pi)^3\delta^{(3)}(p_a-p_b).
\end{equation}
For an $N_{\mathcal F}$-particle state in a box of volume $V$, the diagonal normalization factor is
\begin{equation}
    \mathcal N_{\mathcal F}
    =
    \prod_{r\in\mathcal F} 2E_r V .
\end{equation}
The box regularization replaces the diagonal delta-function singularities by
\begin{equation}
    \tau\equiv 2\pi\delta(0),
    \qquad
    V\equiv (2\pi)^3\delta^{(3)}(0),
\end{equation}
or equivalently,
\begin{equation}
    \left[(2\pi)^4\delta^{(4)}
    (p^\tot_{\mathcal I}-p^\tot_{\mathcal F})\right]^2
    \rightarrow
    \tau V\,
    (2\pi)^4\delta^{(4)}
    (p^\tot_{\mathcal I}-p^\tot_{\mathcal F}) .
\end{equation}
Here $\tau$ is the finite observation or coarse-graining time.

All leading-order formulae below are understood in the perturbative event-probability regime,
\begin{equation}
    P_{\rm event}\ll 1 .
\end{equation}
For a decay, this reduces to $P_{\rm event}=\Gamma\tau\ll1$. For scattering, $P_{\rm event}$ is the corresponding per-event or per-pair transition probability. When the same construction is embedded into a kinetic description, $\tau$ should also lie in the usual coarse-graining window
\begin{equation}
    \tau_{\rm micro}\ll \tau\ll \tau_{\rm relax},
    \label{eq:kinetic-regime}
\end{equation}
where $\tau_{\rm micro}$ is the microscopic duration of a local collision or decay process, while $\tau_{\rm relax}$ is the macroscopic time scale over which distribution functions or background parameters vary appreciably~\cite{DeGroot:1980dk,Kolb:1990vq}.

For an outgoing density matrix
\begin{equation}
    \rho_{\rm out}
    =
    \mathcal S\rho_{\rm in}\mathcal S^\dagger,
\end{equation}
the reduced flavor density matrix is obtained by tracing over kinematics,
\begin{equation}
    \rho_f
    =
    \Tr_{\rm kin} \left[\rho_{\rm out}\right].
\end{equation}
We define the flavor--kinetic linear entanglement entropy by
\begin{equation}
    \mathcal E_{\rm fk}
    =
    1-\Tr_f \left[\rho_f^2 \right].
    \label{eq:fk-linear-entropy}
\end{equation}
This quantity vanishes for a product state between flavor and kinematics and becomes nonzero when the scattering or decay process correlates the discrete branch label with unresolved kinematic configurations. We adopt the linear entropy of Eq.~\eqref{eq:fk-linear-entropy}, rather than the von Neumann entropy, as our entanglement measure throughout, and refer to it simply as the entanglement entropy in what follows.

\subsection{General event-level form}
\label{subsec:general-event-form}

We now derive the leading flavor--kinetic entanglement entropy for a general transition out of an initial branch $\mathcal I$. The derivation applies to both one-particle decays and scattering processes. We write
\begin{equation}
    \mathcal S=1+i\mathcal T .
\end{equation}
The initial branch contains $N_{\mathcal I}$ particles with momenta $p_{\mathcal I}=\{p_r\}_{r\in\mathcal I}$ and total four-momentum
\begin{equation}
    p^\tot_{\mathcal I}
    =
    \sum_{r\in\mathcal I}p_r .
\end{equation}
The normalized incoming density matrix is
\begin{equation}
    \rho_{\rm in}
    =
    \frac{1}{\mathcal N_{\mathcal I}}
    \ket{X_{\mathcal I};p_{\mathcal I}}
    \bra{X_{\mathcal I};p_{\mathcal I}} .
\end{equation}

For a final flavor branch $\mathcal F$, with arbitrary allowed multiplicity, we define the invariant amplitude by
\begin{align}
    \bra{X_{\mathcal F};p_{\mathcal F}}
    \mathcal T
    \ket{X_{\mathcal I};p_{\mathcal I}}
    =
    (2\pi)^4
    \delta^{(4)}(p^\tot_{\mathcal I}-p^\tot_{\mathcal F})
    \mathcal M_{\mathcal F,\mathcal I}
    (p_{\mathcal F},p_{\mathcal I}) .
    \label{eq:general-transition-amplitude}
\end{align}
The Lorentz-invariant phase-space measure is
\begin{align}
    \dd\Phi_{\mathcal F}(p)
    &=
    \frac{1}{S_{\mathcal F}}
    \left[\prod_{r\in\mathcal F}\dd\Pi_r\right]
    (2\pi)^4
    \delta^{(4)}
    \!\left(p-\sum_{r\in\mathcal F}p_r\right),
    \label{eq:multi-particle-phase-space}
    \\
    \dd\Pi_r
    &=
    \frac{\dd^3p_r}{(2\pi)^3 2E_r}.
\end{align}
Here $S_{\mathcal F}$ is the identical-particle symmetry factor of the final branch, equal to the product of factorials for identical particles in $\mathcal F$.

With the covariant box normalization above, the diagonal transition probability from $\mathcal I$ to $\mathcal F$ is
\begin{equation}
    P_{\mathcal F,\mathcal I}
    =
    \frac{\tau}
    {
    V^{N_{\mathcal I}-1}
    \prod_{r\in\mathcal I}2E_r
    }
    \int \dd\Phi_{\mathcal F}(p^\tot_{\mathcal I})
    \left|
    \mathcal M_{\mathcal F,\mathcal I}
    \right|^2 .
    \label{eq:event-branch-probability}
\end{equation}
The symmetry factor in $\dd\Phi_{\mathcal F}$ removes the overcounting of unordered identical final particles. No analogous initial-state symmetry factor appears in this event-level probability, because the incoming microscopic branch is fixed rather than integrated over.

For the flavor--kinetic bipartition, the relevant summed probability is not necessarily the total scattering probability. It is the probability to populate a flavor branch different from the incoming one. We denote the set of resolved branch-changing channels by
\begin{equation}
    \mathcal{BC}(\mathcal I)
    =
    \left\{
    \mathcal F\,\middle|\,\mathcal F\not\simeq \mathcal I
    \right\},
\end{equation}
where $\mathcal F\simeq\mathcal I$ means the same discrete flavor branch, irrespective of the final kinetic configuration. The total branch-changing probability is
\begin{equation}
    P_{\mathcal{BC}(\mathcal I)}
    =
    \sum_{\mathcal F\in\mathcal{BC}(\mathcal I)}
    P_{\mathcal F,\mathcal I}
    \ll1 .
    \label{eq:total-inclusive-event-probability}
\end{equation}
For a one-particle decay, $P_{\mathcal{BC}(\mathcal I)}$ is the total decay probability. For a scattering process, it is the probability summed over all resolved final flavor branches different from the incoming branch. Same-branch (flavor-preserving) transitions, which merely redistribute momenta, remain in the $\mathcal I$ flavor block and do not by themselves generate flavor--kinetic entanglement at leading order.

After tracing over the kinematic degrees of freedom, the reduced flavor density matrix takes the block form
\begin{align}
    \rho_f
    =&\ket{X_{\mathcal I}}\bra{X_{\mathcal I}}
    +\sum_{\mathcal F}\left(B_{\mathcal I\mathcal F}\ket{X_{\mathcal I}}\bra{X_{\mathcal F}}+\mathrm{h.c.}\right)
    \notag\\
    &+\sum_{\mathcal F,\mathcal F'}C_{\mathcal F\mathcal F'}\ket{X_{\mathcal F}}\bra{X_{\mathcal F'}}
    \notag\\
    =&\left(1-P_{\mathcal{BC}(\mathcal I)}\right)    \ket{X_{\mathcal I}}\bra{X_{\mathcal I}}
    \notag\\
    &+\sum_{\mathcal F\in\mathcal{BC}(\mathcal I)}    \left(B_{\mathcal I\mathcal F}    \ket{X_{\mathcal I}}\bra{X_{\mathcal F}}+\mathrm{h.c.}    \right)
    \notag\\
    &+\sum_{\mathcal F,\mathcal F'\in\mathcal{BC}(\mathcal I)}
    C_{\mathcal F\mathcal F'}\ket{X_{\mathcal F}}\bra{X_{\mathcal F'}} ,
    \label{eq:general-flavor-block}
\end{align}
where the incoming--final flavor-coherence block is
\begin{align}
    B_{\mathcal I\mathcal F}
    =
    \frac{\tau}
    {
    V^{N_{\mathcal I}-1}
    \prod_{r\in\mathcal I}2E_r
    }
    \int
    \dd\Phi_{\mathcal I\mathcal F}(p^\tot_{\mathcal I})
    \,
    i\mathcal M_{\mathcal F,\mathcal I}(p_{\mathcal F},p_{\mathcal I}) ,
    \label{eq:incoming-final-coherence-block}
\end{align}
and the final--final flavor-coherence block is
\begin{align}
    C_{\mathcal F\mathcal F'}
    &=
    \frac{\tau}
    {
    V^{N_{\mathcal I}-1}
    \prod_{r\in\mathcal I}2E_r
    }
    \notag\\
    &\quad\times
    \int
    \dd\Phi_{\mathcal F\mathcal F'}(p^\tot_{\mathcal I})
    \,
    \mathcal M_{\mathcal F,\mathcal I}(q,p_{\mathcal I})\,
    \mathcal M_{\mathcal F',\mathcal I}^{*}(q,p_{\mathcal I}) .
    \label{eq:final-final-coherence-block}
\end{align}
In the last line of Eq.~\eqref{eq:general-flavor-block}, we use the optical theorem to replace the coefficient of the term $\ket{X_{\mathcal I}}\bra{X_{\mathcal I}}$ with $1-P_{\mathcal{BC}(\mathcal I)}$.
Both coherence blocks are integrated over the kinetic configuration shared by the two branches they connect, denoted collectively by $q=\{q_r\}$.
The common phase-space measure, written $\dd\Phi_{\mathcal I\mathcal F}$ and $\dd\Phi_{\mathcal F\mathcal F'}$ above, is defined by
\begin{align}
    \dd\Phi_{\mathcal F\mathcal F'}(p)
    &=
    \frac{\delta(m_\mathcal F,m_{\mathcal F'})}
    {\sqrt{S_{\mathcal F}S_{\mathcal F'}}}
    \left[
    \prod_{r=1}^{N_{\mathcal F}}
    \frac{\dd^3q_r}{(2\pi)^3 2E_r(q_r)}
    \right]
    \notag\\
    &\quad\times
    (2\pi)^4
    \delta^{(4)}
    \!\left(p-\sum_{r=1}^{N_{\mathcal F}}q_r\right).
    \label{eq:common-phase-space}
\end{align}
Here $\delta(m_\mathcal F,m_{\mathcal F'})$ equals one only when $\mathcal F$ and $\mathcal F'$ have the same particle number and the same on-shell mass multiset, so that their kinetic configurations can be identified in the trace, and it vanishes otherwise. For $\mathcal F'=\mathcal F$, Eq.~\eqref{eq:common-phase-space} reduces to Eq.~\eqref{eq:multi-particle-phase-space}, and therefore
\begin{equation}
    C_{\mathcal F\mathcal F}
    =
    P_{\mathcal F,\mathcal I}.
\end{equation}
The off-diagonal entries $C_{\mathcal F\mathcal F'}$ encode coherent information among final flavor branches with common kinematic support. 

With the reduced density matrix in Eq.~\eqref{eq:general-flavor-block}, the event-level flavor--kinetic entanglement entropy is
\begin{equation}
\mathcal E_{\rm fk}(\rho_{\rm in})
=
1-\Tr_f\rho_f^2
=
2P_{\mathcal{BC}(\mathcal I)}
+
O\left(P_{\mathcal{BC}(\mathcal I)}^2\right).
\label{eq:general-inclusive-relation}
\end{equation}
Here the off-diagonal incoming--final coherence encoded in $B_{\mathcal I\mathcal F}$ and the resolved final-state block $C_{\mathcal F\mathcal{F}'}$ enter the purity only through quadratic combinations and therefore contribute only at $O(P_{\mathcal{BC}(\mathcal I)}^2)$.

The diagonal part of $\mathcal S\rho_{\rm in}\mathcal S^\dagger$ gives the resolved branch-changing probabilities, while unitarity fixes the corresponding depletion of the incoming flavor block. Final-state flavor coherences affect the detailed reduced density matrix, but they enter the linear entropy only at quadratic order in the small event probabilities. This is why the linear entropy is especially useful for the probability dictionary developed below. By contrast, the von Neumann entropy resolves the small eigenvalues of the final-branch block and contains the nonanalytic dependence $-\lambda\ln\lambda$, making it less directly expressible as a single summed transition probability.

\subsection{Leading-order relations for decay and scattering}
\label{subsec:inclusive-relations}

The general relation in Eq.~\eqref{eq:general-inclusive-relation} becomes concrete once the initial branch is specified. For decay, every genuine decay channel belongs to a flavor branch orthogonal to the one-particle incoming branch. For scattering, by contrast, final states that preserve the full discrete branch remain in the same reduced flavor block and must be excluded from the leading flavor--kinetic entanglement entropy. Thus, the probability entering Eq.~\eqref{eq:general-inclusive-relation} is always the branch-changing probability selected by the flavor--kinetic bipartition.

\paragraph{Decay.}

Consider a one-particle incoming branch $X_{\mathcal I}$ with four-momentum $p_{\mathcal I}$. The allowed decay products can have any final multiplicity $n\ge2$,
\begin{equation}
    X_{\mathcal I}(p_{\mathcal I})
    \longrightarrow
    X_{\mathcal F_n}(p_{\mathcal F_n}) .
\end{equation}
For a decay process, the final particle-content branch is orthogonal to the incoming one-particle branch. Therefore
\begin{equation}
    P_{\mathcal{BC}(\mathcal I)}
    =
    \tau\,\Gamma_{\rm tot}^{\mathcal I},
    \qquad
    \Gamma_{\rm tot}^{\mathcal I}
    =
    \sum_{n\ge2}
    \sum_{\mathcal F_n}
    \Gamma_{\mathcal I\to\mathcal F_n},
\end{equation}
with
\begin{equation}
    \Gamma_{\mathcal I\to\mathcal F_n}(p_{\mathcal I})
    =
    \frac{1}{2E_{\mathcal I}}
    \int
    \dd\Phi_{\mathcal F_n}(p_{\mathcal I})
    \left|
    \mathcal M_{\mathcal F_n,\mathcal I}
    \right|^2 .
    \label{eq:partial-decay-width}
\end{equation}
Here $\Gamma_{\mathcal I\to\mathcal F_n}(p_{\mathcal I})$ denotes the decay rate with respect to the time variable used in the box regularization. In the rest frame of the decaying particle, $E_{\mathcal I}=m_{\mathcal I}$, and $\Gamma_{\mathcal I\to\mathcal F_n}$ reduces to the usual partial width. The general event-level relation therefore gives
\begin{equation}
    \mathcal E_{\rm fk}(\rho_{\rm in})
    =
    2\Gamma_{\rm tot}^{\mathcal I}\tau
    +
    O\!\left[
    \left(\Gamma_{\rm tot}^{\mathcal I}\tau\right)^2
    \right].
    \label{eq:inclusive-decay-relation}
\end{equation}

\paragraph{Scattering.}

Now consider a two-particle incoming branch
\begin{equation}
    X_a(p_a)+X_b(p_b)
    \longrightarrow
    X_{\mathcal F_n}(p_{\mathcal F_n}),
    \qquad
    n=2,3,\ldots .
\end{equation}
The notation $ab$ labels the discrete two-particle incoming branch. Although the branch label may be treated as an unordered flavor multiset, the microscopic incoming state in the event-level probability is fixed. Therefore no initial-state symmetry factor appears in the event-level transition probability. Identical-particle symmetry factors for final states are already contained in the final phase-space measure $\dd\Phi_{\mathcal F_n}$. Initial-state combinatorial factors enter only later when one averages over an ensemble of unordered incoming pairs, as in reaction densities or Boltzmann collision terms.

For the branch $ab\to\mathcal F_n$, Eq.~\eqref{eq:event-branch-probability} gives
\begin{equation}
    P_{\mathcal F_n,ab}
    =
    \frac{\tau}
    {V\,2E_a\,2E_b}
    \int
    \dd\Phi_{\mathcal F_n}(p_a+p_b)
    \left|
    \mathcal M_{\mathcal F_n,ab}
    \right|^2 .
    \label{eq:scattering-branch-probability}
\end{equation}
Equivalently, using the standard invariant flux factor, define
\begin{equation}
    \sigma_{ab\to\mathcal F_n}
    =
    \frac{1}
    {4E_aE_b v_{\rm rel}}
    \int
    \dd\Phi_{\mathcal F_n}(p_a+p_b)
    \left|
    \mathcal M_{\mathcal F_n,ab}
    \right|^2 ,
\end{equation}
where $v_{\rm rel}$ is the M{\o}ller relative velocity,
\begin{equation}
    v_{\rm rel}
    =
    \frac{
    \sqrt{(p_a\cdot p_b)^2-m_a^2m_b^2}
    }
    {E_aE_b}.
\end{equation}
Then
\begin{equation}
    P_{\mathcal F_n,ab}
    =
    \frac{\tau}{V}
    v_{\rm rel}
    \sigma_{ab\to\mathcal F_n}.
    \label{eq:scattering-probability-cross-section}
\end{equation}

The leading flavor--kinetic entanglement entropy is controlled only by final branch-changing channels. Define the total branch-changing cross section
\begin{equation}
    \sigma_{\mathcal{BC}(ab)}
    =
    \sum_{n\ge2}
    \sum_{\mathcal F_n\in\mathcal{BC}(ab)}
    \sigma_{ab\to\mathcal F_n}.
    \label{eq:branch-changing-cross-section}
\end{equation}
Then
\begin{equation}
    P_{\mathcal{BC}(ab)}
    =
    \frac{\tau}{V}
    v_{\rm rel}
    \sigma_{\mathcal{BC}(ab)},
\end{equation}
and Eq.~\eqref{eq:general-inclusive-relation} gives
\begin{equation}
    \mathcal E_{\rm fk}(\rho_{\rm in})
    =
    2
    \frac{\tau}{V}
    v_{\rm rel}
    \sigma_{\mathcal{BC}(ab)}
    +
    O\!\left(P_{\mathcal{BC}(ab)}^2\right).
    \label{eq:inclusive-scattering-relation}
\end{equation}
This is the scattering analog of the decay relation: the total decay rate is replaced by the total branch-changing cross section selected by the flavor--kinetic bipartition.

\section{From vacuum to finite-density flavor--kinetic entanglement entropy}
\label{sec:finite-density}

This section extends the event-level construction of Sec.~\ref{sec:smatrix} to a finite-density background.  The microscopic input is still an ordinary Fock-space amplitude $\M_{cd,ab}$ between quasiparticle branches. When medium effects are retained, it is evaluated with thermal masses, effective couplings, screening effects, or resummed propagators, and we write it explicitly as $\M^{\mathrm{FD}}_{cd,ab}$.

The finite-density construction in this section is restricted to $2\to2$ scattering channels, since these are the channels used in the thermal application below in Sec.~\ref{sec:thermal-application}.

\subsection{Construction of the finite-density background}
\label{subsec:fd-construction}

The finite-density background supplies incoming occupation functions and final-state statistical factors through a product quasiparticle ensemble,
\begin{align}
\rho_{\rm FD}
&=
\bigotimes_r \rho_r[f_r],
\notag\\
\Tr_{\mathcal H_{\rm Fock}}\rho_{\rm FD}
&=1,
\qquad
\Tr_{\mathcal H_{{\rm Fock},r}}\rho_r[f_r]
=1 .
\end{align}
The one-particle occupation number is fixed by
\begin{align}
&\Tr_{\mathcal H_{{\rm Fock},r}}\!\left[
\rho_r[f_r]\,
a_r^\dagger(\mathbf p)a_r(\mathbf p')
\right]
\notag\\
&\qquad =
(2\pi)^3 2E_r(\mathbf p)\,
\delta^3(\mathbf p-\mathbf p')\,
f_r(\mathbf p).
\end{align}
The number density of species $X_r$ is
\begin{equation}
n_r
=
\int\frac{\dd^3p}{(2\pi)^3}\,
f_r(\mathbf p).
\label{eq:number-density}
\end{equation}
For later use, the effective one-particle density matrix is defined by
\begin{equation}
\rho_r^{(1)}
=
\frac{1}{n_rV}
\int \dd\Pi_r\,
f_r(\mathbf p_r)
\ket{X_r;p_r}\bra{X_r;p_r},
\label{eq:rho-a}
\end{equation}
which cannot represent the most general finite-density background but is valid when the species $r$ can be treated as an on-shell,
homogeneous, and incoherent kinetic ensemble, whose only retained
one-particle information is the classical momentum distribution
$f_r(\mathbf p_r)$. This form is therefore valid
for one-body observables or for scattering from an uncorrelated bath,
while it neglects off-diagonal coherences of the form
$\ket{X_r;p}\bra{X'_r;p'}$ and genuine many-body correlations.  If spin, flavor, or medium-induced coherences are relevant, $f_r(\mathbf p_r)$ should be promoted to a matrix of densities $\varrho_r(\mathbf p_r)$ acting on the internal labels, while coherence between different momentum modes---i.e., spatial inhomogeneity---further requires its Wigner-function generalization $\varrho_r(\mathbf x,\mathbf p_r)$~\cite{Stodolsky:1986dx,Raffelt:1992uj,Sigl:1993ctk}.

For an unordered incoming pair $\{a,b\}$, the $2\to2$ subprocess is
\begin{equation}
X_a(p_a)+X_b(p_b)
\longrightarrow
X_c(p_c)+X_d(p_d).
\end{equation}
In the factorized quasiparticle approximation used below, the diagonal incoming pair weight is $f_a(\mathbf p_a)f_b(\mathbf p_b)$,
with the initial-state identical-particle symmetry factor $S_{ab}=1+\delta_{ab}$. For identical incoming particles the two-particle states are the symmetrized Fock states of Sec.~\ref{subsec:hilbert-space}.
The final-state Bose-enhancement or Pauli-blocking factor is denoted by
\begin{equation}
\Xi_r(\mathbf p_r)
=
1+\eta_r f_r(\mathbf p_r),
\qquad
\eta_r=
\begin{cases}
+1,& X_r\ {\rm bosonic},\\
-1,& X_r\ {\rm fermionic}.
\end{cases}
\label{eq:stat-factor}
\end{equation}
The finite-density transition probability for a single incoming $ab$ pair drawn from the bath, with momenta distributed according to $f_af_b/(n_an_b)$, to scatter into the final branch $cd$ is
\begin{align}
P^{\rm FD}_{cd,ab}
&=
S_{ab}\,\frac{\tau}{n_a n_b V} \Gamma_{ab\to cd},
\label{eq:fd-channel-probability}
\end{align}
where the reaction density $\Gamma_{ab\to cd}$ is
\begin{align}
    \Gamma_{ab\to cd}
    &=
    \frac{1}{S_{ab}}
    \int
    \dd\Pi_a\dd\Pi_b\,
    f_a(\mathbf p_a)f_b(\mathbf p_b)
    \notag\\
    &\quad\times
    \int
    \dd\Phi_{cd}(p_a+p_b)\,
    \Xi_c(\mathbf p_c)\Xi_d(\mathbf p_d)
    \left|
    \M^{\rm FD}_{cd,ab}
    \right|^2 .
    \label{eq:reaction-density}
\end{align}
The probability $P^{\rm FD}_{cd,ab}$ is the finite-density analog of the fixed-momentum event probability in Eq.~\eqref{eq:scattering-branch-probability}, averaged over the incoming thermal momenta and supplemented by the final-state statistical factors. Equivalently, $P^{\rm FD}_{cd,ab}=(\tau/V)\,\langle\sigma^{\rm FD}_{cd}\,v_{\rm rel}\rangle$, where $\langle\sigma^{\rm FD} v_{\rm rel}\rangle\equiv(n_an_b)^{-1}\int\!\frac{\dd^3p_a}{(2\pi)^3}\frac{\dd^3p_b}{(2\pi)^3}\,f_af_b\,\sigma^{\rm FD}v_{\rm rel}$ and $\sigma^{\rm FD}$ is the in-medium cross section including the final-state statistical factors. The factor $S_{ab}$ in Eq.~\eqref{eq:fd-channel-probability} cancels the identical-particle factor $1/S_{ab}$ of Eq.~\eqref{eq:reaction-density}, whose role is to convert the ordered momentum integration into an event count. Equation~\eqref{eq:fd-channel-probability} is therefore the plain $f_af_b/(n_an_b)$ average of the fixed-pair probability of Eq.~\eqref{eq:scattering-branch-probability}: no symmetry factor appears at the event level (cf.\ Sec.~\ref{subsec:general-event-form}). The factors $\Xi_c\Xi_d$ and the dressed amplitude $\M^{\rm FD}$ encode the effect of the spectator bath and are inserted according to the standard finite-density collision-term construction~\cite{DeGroot:1980dk,Kolb:1990vq}. In the present construction the background therefore enters only through the sampling measure and this medium dressing of the collision. The entanglement entropy itself is defined for the sampled pair in the next subsection, not for the full ensemble.

\subsection{Bath-averaged flavor--kinetic entanglement entropy for finite-density backgrounds}

The finite-density generalization is defined per pair sampled from the bath. For an unordered incoming pair $\{a,b\}$, the pair-conditioned incoming state is built from the effective one-particle density matrices of Eq.~\eqref{eq:rho-a},
\begin{align}
\rho^{(ab)}_{\rm in}
=
\rho_a^{(1)}\otimes\rho_b^{(1)},
\label{eq:fd-pair-state}
\end{align}
with the Fock-space symmetrization of Sec.~\ref{subsec:hilbert-space} understood for identical species. The flavor reduction of $\rho^{(ab)}_{\rm in}$ is the pure branch projector $\ket{X_{ab}}\bra{X_{ab}}$: the momentum mixture resides entirely in the kinematic factor, so the flavor--kinetic entanglement entropy vanishes before the collision and is generated entirely by it. The finite-density flavor--kinetic entanglement entropy of the sampled pair is
\begin{align}
\mathcal E_{\rm fk}^{\rm FD,(ab)}
&=
1-\Tr_f\!\left[\big(\rho_f^{(ab)}\big)^2\right],
\notag\\
\rho_f^{(ab)}
&=
\Tr_{\kin}\!\left(\mathcal S\,\rho^{(ab)}_{\rm in}\,\mathcal S^\dagger\right),
\label{eq:fd-pair-entropy}
\end{align}
with the in-medium amplitude and the final-state statistical factors of Eq.~\eqref{eq:reaction-density}. Because $\mathcal S(\cdot)\mathcal S^\dagger$ and $\Tr_{\kin}$ are linear, $\rho_f^{(ab)}$ is the $f_af_b$-weighted average of the event-level reduced flavor matrices of Sec.~\ref{sec:smatrix}, so that to leading order
\begin{align}
\mathcal E_{\rm fk}^{\rm FD,(ab)}
&=
\int\!\frac{\dd^3p_a\,\dd^3p_b}{(2\pi)^6}\,
\frac{f_af_b}{n_an_b}\,
\mathcal E_{\rm fk}\big(\rho_\mathrm{in}^{(ab)}(p_a,p_b)\big)
+O(P^2)
\notag\\
&=
2\,P^{\rm FD}_{\mathcal{BC}(ab)}
+O(P^2)
\notag\\
&=
\frac{2\tau}{V}
\big\langle
\sigma^{\rm FD}_{\mathcal{BC}(ab)}\,v_{\rm rel}
\big\rangle
+O(P^2),
\label{eq:fd-theorem}
\end{align}
where $P^{\rm FD}_{\mathcal{BC}(ab)}\equiv\sum_{cd\in\mathcal{BC}(ab)}P^{\rm FD}_{cd,ab}$, and $\rho_{\rm in}^{(ab)}(p_a,p_b)=\ket{X_{ab};p_ap_b}\bra{X_{ab};p_ap_b}/(4E_aE_bV^2)$ is the normalized fixed-momentum incoming state of Sec.~\ref{subsec:general-event-form}, of which $\rho^{(ab)}_{\rm in}$ is the $f_af_b/(n_an_b)$ average. Each $\mathcal E_{\rm fk}\big(\rho_\mathrm{in}^{(ab)}(p_a,p_b)\big)$ is the genuine event-level entanglement entropy of Sec.~\ref{sec:smatrix}: the finite-density observable is the \emph{bath-averaged flavor--kinetic entanglement entropy} of a sampled pair, and it generalizes the vacuum relation~\eqref{eq:inclusive-scattering-relation} by thermally averaging $v_{\rm rel}\sigma_{\mathcal{BC}}$ with medium statistics~\cite{Low:2024hvn,Low:2024mrk}.
The coherence-block structure behind Eq.~\eqref{eq:fd-theorem} is detailed in Appendix~\ref{app:finite-density-details}.
The total finite-density entanglement entropy used below aggregates the active species pairs with equal weight,
\begin{align}
\mathcal E^{\rm FD}_{\rm fk}
\equiv
\sum_{\{a,b\}}
\mathcal E^{\rm FD,(ab)}_{\rm fk},
\label{eq:fd-aggregate}
\end{align}
a channel-resolved diagnostic functional. 
It is not the entanglement entropy of the full mixed finite-density ensemble and is not a thermodynamic entropy. 

Within the event-level normalization of Eq.~\eqref{eq:fd-channel-probability}, the leading branch-changing channel contribution is
\begin{equation}
    \mathcal E_{{\rm fk},cd}^{\rm FD,(ab)}
    =
    \frac{2S_{ab}\,\tau}{n_an_bV}\,
    \Gamma_{ab\to cd}
    +O\!\left(P^2\right).
    \label{eq:fd-channel-entropy-kernel}
\end{equation}
This positive channel contribution will be recast below as a coarse-grained entropy rate in the integrated Boltzmann equation.

\subsection{Relation to integrated Boltzmann equations}
\label{subsec:boltzmann-kernel}

The same finite-density reaction density in Eq.~\eqref{eq:reaction-density} appears in the integrated Boltzmann equation.  With the same unordered-pair convention, one may write
\begin{equation}
    \dot n_a+3Hn_a
    =
    \sum_{b,\{cd\}}
    \Delta N_a^{ab\to cd}
    \left[
    \Gamma_{ab\to cd}
    -
    \Gamma_{cd\to ab}
    \right],
    \label{eq:boltzmann-stoichiometric}
\end{equation}
where $\Delta N_a^{ab\to cd}$ is the net number of particles $a$ produced in one $ab\to cd$ event, and $H$ is the Hubble rate.  Detailed balance rewrites each gain-minus-loss bracket as a departure from equilibrium multiplied by the same positive channel kernel, without introducing any additional collision object.
At leading order, one can define the rate of increase of coarse-grained entanglement by $\dot{\mathcal E}_{{\rm fk},cd}^{\rm FD,(ab)}\equiv \mathcal E_{{\rm fk},cd}^{\rm FD,(ab)}/\tau$. Since branch-preserving channels carry $\Delta N_a=0$ and drop out of Eq.~\eqref{eq:boltzmann-stoichiometric} identically, restricting the sum to the branch-changing set $\mathcal{BC}(ab)$ only assumes that the retained $2\to2$ channels exhaust the number-changing reactions of $X_a$. With this restriction,
\begin{equation}
\begin{split}
    \dot n_a+3Hn_a
    &\simeq
    \frac{V}{2}
    \sum_{b,\{cd\}\in\mathcal{BC}(ab)}
    \Delta N_a^{ab\to cd}
    \\
    &\quad\times
    \left[
    \frac{n_an_b}{S_{ab}}\,\dot{\mathcal E}_{{\rm fk},cd}^{\rm FD,(ab)}
    -
    \frac{n_cn_d}{S_{cd}}\,\dot{\mathcal E}_{{\rm fk},ab}^{\rm FD,(cd)}
    \right].
\end{split}
\label{eq:boltzmann-directed-entropy-kernels}
\end{equation}
Note that $\mathcal{BC}(ab)$ may itself contain channels with $\Delta N_a=0$, such as spectator-type reactions $ab\to ad$, so the uniformity condition below is an additional restriction rather than a consequence of the branch-changing selection.
If $\Delta N_a^{ab\to cd}=\Delta N_a^{ab}$ for every included channel, then in the kinetic-equilibrium, dilute Maxwell--Boltzmann limit, $f_i=z_i f_i^{\rm eq}$, with microscopic reversibility, $z_cz_d=1$, and equilibrium-fixed medium parameters, detailed balance yields
\begin{equation}
\begin{split}
    \dot n_a+3Hn_a
    &\simeq
    \frac{V}{2}
    \sum_b
    \frac{\Delta N_a^{ab}}{S_{ab}}
    \\
    &\quad\times
    \left(n_an_b-n_a^{\rm eq}n_b^{\rm eq}\right)
    \dot{\mathcal E}_{{\rm fk}}^{\rm FD,(ab)}.
\end{split}
\label{eq:boltzmann-equilibrium-entropy-kernel}
\end{equation}
The signed factor $\Delta N_a^{ab}$ encodes depletion, while $\dot{\mathcal E}_{\rm fk}^{\rm FD,(ab)}$ measures the coarse-grained per-pair branch-changing collision rate, $2\langle\sigma^{\rm FD}_{\mathcal{BC}}v_{\rm rel}\rangle/V$, not thermodynamic entropy production.  It remains positive at equilibrium although the net abundance flow vanishes, which sets chemical relaxation.  Its thermal-background dependence can therefore diagnose phase-transition-type behavior.

\section{Application: finite-density flavor--kinetic entanglement entropy as a phase-transition-type diagnostic}
\label{sec:thermal-application}

\begin{figure*}[t]
\centering
\begin{minipage}{0.32\textwidth}
\centering
\includegraphics[width=\textwidth]{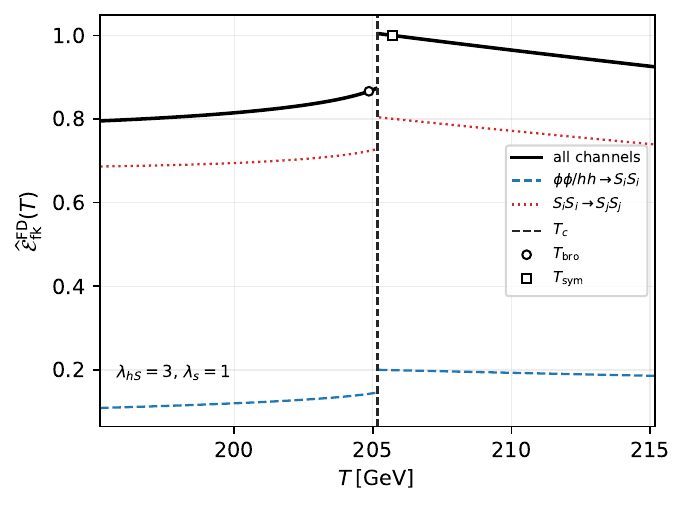}\\
(a) FOPT
\end{minipage}
\hfill
\begin{minipage}{0.32\textwidth}
\centering
\includegraphics[width=\textwidth]{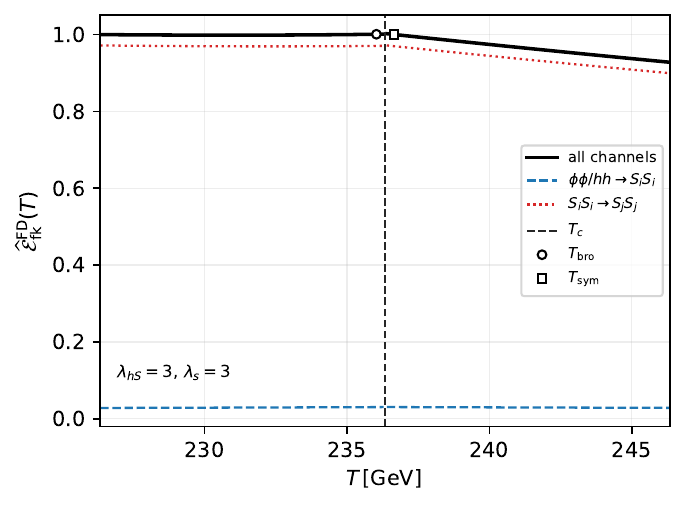}\\
(b) Continuous transition
\end{minipage}
\hfill
\begin{minipage}{0.32\textwidth}
\centering
\includegraphics[width=\textwidth]{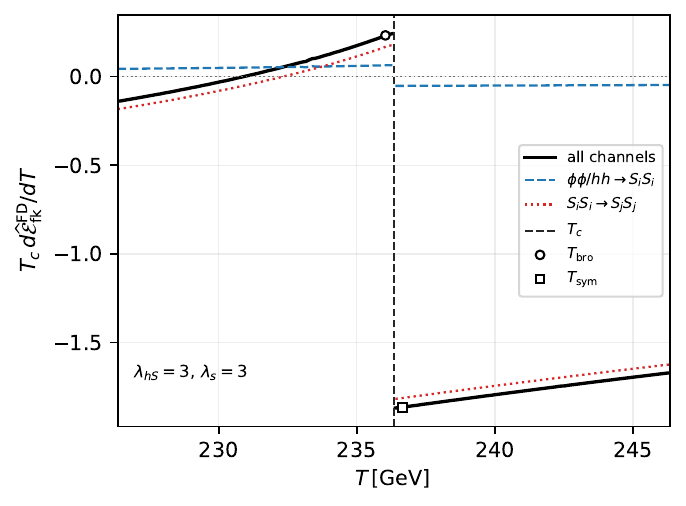}\\
(c) Derivative near a continuous transition
\end{minipage}
\caption{Benchmark finite-density flavor--kinetic entanglement entropy. Panel (a) shows the first-order benchmark, where $\widehat{\mathcal E}_{\rm fk}(T)$ jumps across $T_c$. Panel (b) shows the continuous benchmark, where $\widehat{\mathcal E}_{\rm fk}(T)$ itself remains continuous. Panel (c) shows the corresponding nonanalytic behavior in the first temperature derivative. Dashed and dotted curves are channel-family components of the same entanglement-entropy sum, not independent entropy definitions.}
\label{fig:benchmark-response}
\end{figure*}

This section applies the finite-density flavor--kinetic entanglement entropy defined in Sec.~\ref{sec:finite-density} to a thermal bath. In the examples below, an FOPT can produce a finite discontinuity in this entanglement entropy, while a continuous transition can leave the entanglement entropy itself continuous but generate a nonanalytic temperature derivative.  We use this behavior as a phase-transition-type diagnostic.  

The perturbative finite-temperature effective potential is used here only as a standard phenomenological input for phase structure. This input inherits the usual gauge, infrared-resummation, and renormalization-scheme uncertainties of perturbative thermal effective-potential calculations~\cite{Dolan:1973qd,Carrington:1991hz,Quiros:1999jp,Ekstedt:2020abj}.  The potential input, the thermal self-energies, and the numerical procedure used below are collected in Appendix~\ref{app:on-numerics}.

\subsection{Thermal branch-changing entanglement entropy}
\label{subsec:thermal-response-definition}

We evaluate the finite-density aggregate in Eq.~\eqref{eq:fd-aggregate} on a thermal quasiparticle background, with distribution functions $f_r(\mathbf p;T)$, thermal masses, and medium-dependent amplitudes evaluated at the same temperature.  For compact notation, we write $\mathcal E_{\rm fk}^{\rm FD}(T)$ for this aggregate evaluated at temperature $T$.

We compare the flavor--kinetic entanglement entropy just above and just below the critical temperature 
\begin{equation}
T_{\rm sym}=T_c+\delta T,
\qquad
T_{\rm bro}=T_c-\delta T .
\end{equation}
Throughout the numerical evaluation we use the fixed offset $\delta T=0.31~{\rm GeV}$, set by the temperature grid on which the thermal background and the reaction densities are computed.  Since $\widehat{\mathcal E}_{\rm fk}(T)$ is continuous across a continuous transition, the residual $\mathcal J_{\rm fk}$ measured there is of order $\delta T$ and vanishes as $\delta T\to0$, whereas across a first-order transition $\mathcal J_{\rm fk}$ approaches a finite $\delta T$-independent limit.  The contour level used in the scans below is chosen against this residual.
We then define the dimensionless finite-density flavor--kinetic entanglement entropy,
\begin{equation}
\widehat{\mathcal E}_{\rm fk}(T)
=
\frac{\mathcal E^\mathrm{FD}_{\rm fk}(T)}
{\mathcal E^\mathrm{FD}_{\rm fk}(T_{\rm sym})}.
\label{eq:Ehat_fk}
\end{equation}
The common regularization factor $\tau/V$ cancels in the normalized quantities.
The signed dimensionless jump is
\begin{equation}
\mathcal J_{\rm fk}
=
\widehat{\mathcal E}_{\rm fk}(T_{\rm bro})-1.
\label{eq:J_fk}
\end{equation}
A positive $\mathcal J_{\rm fk}$ means that the flavor branch-changing collision kernel becomes more efficient in the broken phase, while a negative value means that it becomes less efficient.  The magnitude $|\mathcal J_{\rm fk}|$ measures the size of the discontinuity in the entanglement entropy.  A value consistent with zero means only that $\widehat{\mathcal E}_{\rm fk}(T)$ is smooth at the chosen numerical resolution.
For continuous transitions the diagnostic is carried by the first temperature derivative, for which we use the dimensionless rescaling
\begin{equation}
\widehat{\mathcal E}'_{\rm fk}(T)
\equiv
T_c\,\frac{\dd\widehat{\mathcal E}_{\rm fk}(T)}{\dd T} .
\label{eq:Ehat_fk_derivative}
\end{equation}

\subsection{\texorpdfstring{Minimal $O(N)$}{Minimal O(N)} scalar extended model}
\label{subsec:on-model}

\begin{figure*}[t]
\centering
\begin{minipage}{0.32\textwidth}
\centering
\includegraphics[width=\textwidth]{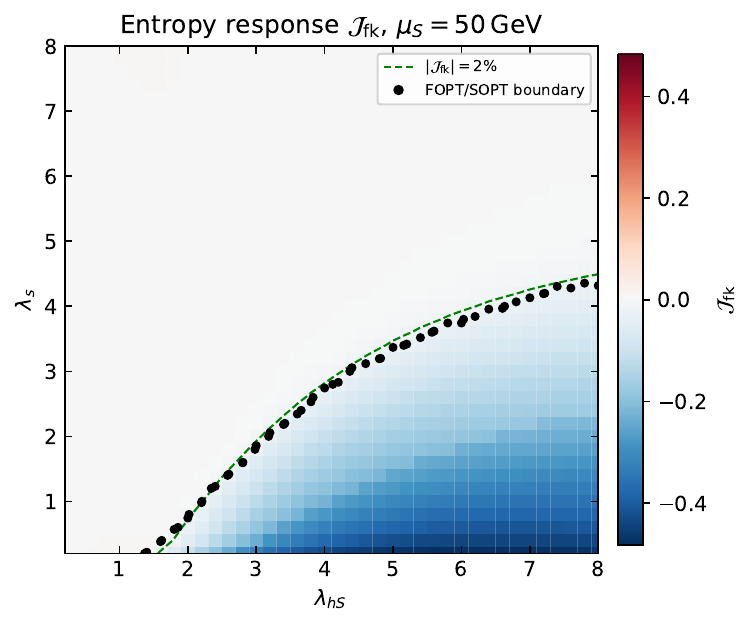}\\
(a) $\mu_S=50~{\rm GeV}$
\end{minipage}
\hfill
\begin{minipage}{0.32\textwidth}
\centering
\includegraphics[width=\textwidth]{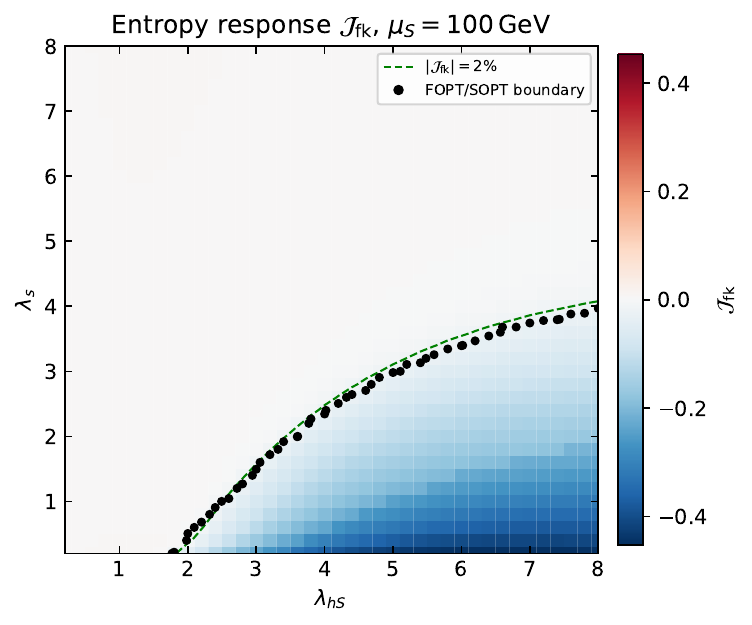}\\
(b) $\mu_S=100~{\rm GeV}$
\end{minipage}
\hfill
\begin{minipage}{0.32\textwidth}
\centering
\includegraphics[width=\textwidth]{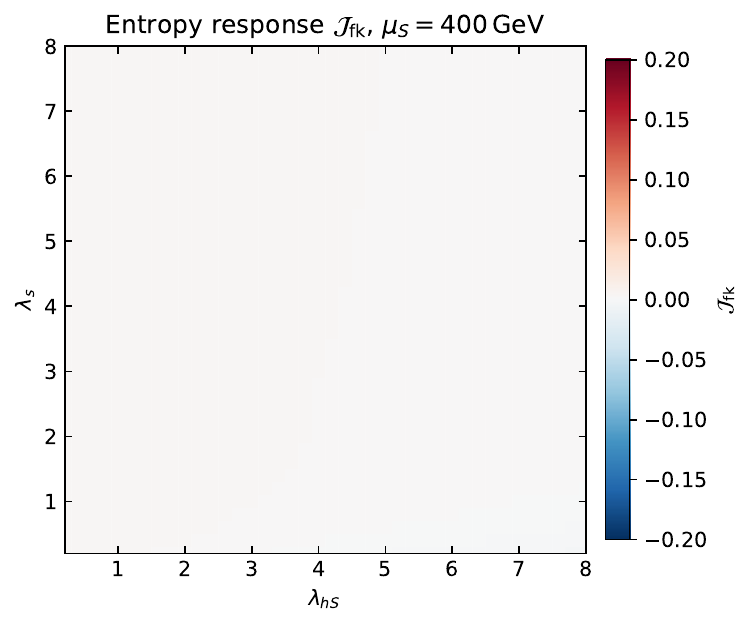}\\
(c) $\mu_S=400~{\rm GeV}$
\end{minipage}
\caption{Parameter scans of the signed dimensionless jump $\mathcal J_{\rm fk}$. Panels (a) and (b) show light-singlet scans for $\mu_S=50~{\rm GeV}$ and $100~{\rm GeV}$. The color maps show $\mathcal J_{\rm fk}=\widehat{\mathcal E}_{\rm fk}(T_{\rm bro})-1$, and green dashed contours mark $|\mathcal J_{\rm fk}|=0.02$ when that level lies in the plotted range. Black markers denote the independently computed boundary between first-order and continuous transitions from the finite-temperature effective potential. Panel (c) shows the heavy-singlet slice for $\mu_S=400~{\rm GeV}$. There the potential scan finds no FOPT region in the plotted domain, and the signed jump $\mathcal J_{\rm fk}$ remains close to zero throughout.}
\label{fig:entropy-response-scans}
\end{figure*}

Here we use a real scalar $\phi$ coupled to an $O(N)$-symmetric multiplet $\vec S=(S_1,\ldots,S_N)$,
\begin{equation}
V(\phi,\vec S)
=
-\frac{\mu_h^2}{2}\phi^2
+
\frac{\lambda_h}{24}\phi^4
+
\frac{\lambda_{hS}}{4}\phi^2\vec S^2
+
\frac{\mu_S^2}{2}\vec S^2
+
\frac{\lambda_S}{4}(\vec S^2)^2 .
\label{eq:on-potential}
\end{equation}
Singlet and Higgs-portal scalar sectors are widely used benchmark settings for electroweak phase-transition studies~\cite{Chung:2012vg,Kurup:2017dzf}. The present application follows the $O(N)$ flavor structure used in Ref.~\cite{Liu:2025pny}, but evaluates the finite-density collision kernel defined in Sec.~\ref{sec:finite-density}.  The scan label $\mu_S$ denotes the bare singlet mass parameter appearing in Eq.~\eqref{eq:on-potential}.

The flavor branch-changing channel families are phase dependent.  In the symmetric phase, the scalar excitation is denoted by $\phi$, and we include
\begin{equation}
\phi\phi\leftrightarrow S_iS_i,
\qquad
S_iS_i\to S_jS_j,
\qquad
i,j=1,\ldots,N,\quad i\neq j .
\label{eq:on-symmetric-channels}
\end{equation}
In the broken phase, the radial excitation is denoted by $h$, and the corresponding channel families are
\begin{equation}
hh\leftrightarrow S_iS_i,
\qquad
S_iS_i\to S_jS_j,
\qquad
i,j=1,\ldots,N,\quad i\neq j .
\label{eq:on-broken-channels}
\end{equation}
These channel families are components of the same finite-density flavor--kinetic entanglement-entropy functional, not separate entropy definitions.

Using the explicit microscopic flavor sums, the symmetric-phase entanglement entropy is
\begin{align}
\mathcal E_{\rm fk}^{\rm FD,sym}(T)
&=
2
\sum_{i=1}^{N}
P^{\rm FD}_{S_iS_i,\phi\phi}(T)+
2\sum_{i=1}^{N}P^{\rm FD}_{\phi\phi,S_iS_i}(T)
\notag\\
&\quad+
2
\sum_{i=1}^{N}
\sum_{\substack{j=1\\ j\neq i}}^{N}
P^{\rm FD}_{S_jS_j,S_iS_i}(T)
+
O(P^2).
\label{eq:on-active-symmetric}
\end{align}
Similarly, the broken-phase entanglement entropy is
\begin{align}
\mathcal E_{\rm fk}^{\rm FD,bro}(T)
&=
2
\sum_{i=1}^{N}
P^{\rm FD}_{S_iS_i,hh}(T)+
2\sum_{i=1}^{N}
P^{\rm FD}_{hh,S_iS_i}(T)
\notag\\
&\quad+
2
\sum_{i=1}^{N}
\sum_{\substack{j=1\\ j\neq i}}^{N}
P^{\rm FD}_{S_jS_j,S_iS_i}(T)
+
O(P^2).
\label{eq:on-active-broken}
\end{align}
The finite-density entanglement entropy used in Eq.~\eqref{eq:Ehat_fk} is then evaluated as
\begin{equation}
\mathcal E_{\rm fk}^{\rm FD}(T)
=
\begin{cases}
\mathcal E_{\rm fk}^{\rm FD,sym}(T), & T>T_c,\\
\mathcal E_{\rm fk}^{\rm FD,bro}(T), & T<T_c.
\end{cases}
\label{eq:on-phase-response}
\end{equation}
The sums over $i$ and $j$ are explicit sums over microscopic $O(N)$ flavor branches.  The families $\phi\phi\leftrightarrow S_iS_i$ or $hh\leftrightarrow S_iS_i$ contain $N$ microscopic channels in each direction, while $S_iS_i\to S_jS_j$ with $i\neq j$ contains $N(N-1)$ ordered flavor-branch transitions.  In the numerical evaluation, the $O(N)$ symmetry makes the members of each channel family degenerate, so Eqs.~\eqref{eq:on-active-symmetric} and~\eqref{eq:on-active-broken} may be evaluated by computing one representative channel in each family and multiplying by the corresponding number of explicit terms.

\subsection{Benchmark points and parameter scans}
\label{subsec:benchmarks-and-scans}
\label{subsec:benchmarks}
\label{subsec:scan}

We first illustrate the finite-density flavor--kinetic entanglement entropy with two $N=4$ benchmark points and then scan the same quantity over the $(\lambda_{hS},\lambda_S)$ plane.  The benchmarks show the two characteristic possibilities relevant for the phase-transition diagnosis: $\widehat{\mathcal E}_{\rm fk}(T)$ can develop a finite jump across an FOPT, while it can remain continuous across a continuous transition with a nonanalytic temperature derivative.  The scan then compares the signed jump $\mathcal J_{\rm fk}$ with the phase classification obtained independently from the finite-temperature effective potential.

The first benchmark has a first-order transition with $\lambda_{hS}=3.0$ and $\lambda_S=1.0$.  As shown in Fig.~\ref{fig:benchmark-response}(a), $\widehat{\mathcal E}_{\rm fk}(T)$ develops a finite jump across $T_c$.  This behavior is the finite-density flavor--kinetic analog of the standard thermodynamic expectation that the type of phase transition is reflected in the nonanalytic structure of a temperature-dependent quantity.
The second benchmark keeps the same portal coupling but raises the singlet self-coupling, $\lambda_{hS}=3.0$ and $\lambda_S=3.0$, and the effective-potential scan classifies its transition as continuous.  As shown in Fig.~\ref{fig:benchmark-response}(b), $\widehat{\mathcal E}_{\rm fk}(T)$ then remains continuous across $T_c$, while its first temperature derivative develops the nonanalytic behavior shown in Fig.~\ref{fig:benchmark-response}(c).
The two-dimensional scans show maps of the signed dimensionless jump $\mathcal J_{\rm fk}$ defined in Eq.~\eqref{eq:J_fk}.  The black dots or black curves overlaid on the density plots are obtained independently from the finite-temperature effective potential.  The flavor--kinetic entanglement entropy is evaluated afterward and compared with this thermal-potential classification.

The light-singlet scans in Figs.~\ref{fig:entropy-response-scans}(a) and~\ref{fig:entropy-response-scans}(b) use $\mu_S=50~{\rm GeV}$ and $100~{\rm GeV}$, respectively.  In the FOPT region, $\widehat{\mathcal E}_{\rm fk}(T)$ develops a finite jump near the critical temperature.  Outside the FOPT region, no finite jump is expected in $\mathcal J_{\rm fk}$.  For continuous-transition points, $\widehat{\mathcal E}_{\rm fk}(T)$ can remain continuous while its first temperature derivative is nonanalytic, as illustrated in Fig.~\ref{fig:benchmark-response}(c).  In the heavier-singlet scan shown in Fig.~\ref{fig:entropy-response-scans}(c), the finite-temperature effective-potential scan finds no FOPT region in the displayed coupling domain, and $\mathcal J_{\rm fk}$ remains close to zero.  This is consistent with the potential-based classification.
In every FOPT region found here $\mathcal J_{\rm fk}$ is negative, reaching $\mathcal J_{\rm fk}\simeq-0.13$ at the first-order benchmark, so that in these examples the branch-changing collisions are less efficient per pair in the broken phase.

The observed correspondence has a natural correlated origin.  The reaction densities in Eq.~\eqref{eq:reaction-density} are evaluated on the equilibrium background $v(T)$: this background controls the dressed quasiparticle masses, phase-space and statistical weights, and the background-expanded amplitudes, including the broken-phase $hS_iS_i$ interaction.  At a first-order transition, a discontinuity of $v(T)$ can therefore induce a discontinuity in the branch-changing collision kernel and hence in $\widehat{\mathcal E}_{\rm fk}(T)$, whereas at a continuous transition these inputs can remain continuous while their temperature derivatives are nonanalytic.  The correlation thus reflects the background dependence of the kernel as a whole rather than any single ingredient.

The application shows that $\mathcal E_{\rm fk}^{\rm FD}(T)$ can serve as a phase-transition-type diagnostic when its temperature dependence develops the nonanalytic structure expected from the thermodynamic classification.  The comparison with the finite-temperature effective potential should therefore be understood as a consistency check: the flavor--kinetic entanglement entropy tracks the same transition pattern in the examples considered, while remaining a collision-kernel-based quantity rather than a thermodynamic observable.

\section{Conclusions}
\label{sec:conclusions}

In this work we extended the scattering--entanglement dictionary to
finite-density settings for the flavor--kinetic bipartition.  For an
isolated event, tracing over the kinematic degrees of freedom gives the
leading-order relation
\begin{equation}
    \mathcal E_{\rm fk}
    =
    2P_{\mathcal{BC}}
    +
    O(P_{\mathcal{BC}}^2),
\end{equation}
where $P_{\mathcal{BC}}$ is the probability for populating final
flavor branches different from the incoming branch.  Thus, same-branch
kinematic scattering does not contribute to the leading flavor--kinetic
entanglement entropy, while incoming--final and final--final flavor coherences
enter the entanglement entropy only at quadratic order in the transition
probabilities.  This makes the leading entanglement entropy a branch-changing
probability selected by the flavor partition.

We then formulated the corresponding finite-density entanglement entropy for the
$2\to2$ quasiparticle channels considered in this work.  The vacuum
branch-changing probability is replaced by the event-level probability
\begin{equation}
    P^{\rm FD}_{cd,ab}
    =
    S_{ab}\,\frac{\tau}{n_a n_b V}\,
    \Gamma_{ab\to cd},
\end{equation}
where $\Gamma_{ab\to cd}$ contains the incoming occupation weights,
the final-state statistical factors, and the finite-density
quasiparticle amplitude.  The same directed reaction density
appears in the integrated Boltzmann equation, but with a different
projection: Boltzmann evolution forms gain-minus-loss combinations
weighted by stoichiometric factors, whereas
$\mathcal E_{\rm fk}^{\rm FD}$ keeps the positive branch-changing
kernels selected by the flavor--kinetic bipartition and normalizes them
per sampled pair.  At leading order this quantity is the bath average of event-level entanglement entropies, aggregated over the active species pairs with equal weight.  This bath-average interpretation is what justifies reading it as finite-density entanglement production.  In this sense, the construction connects
microscopic entanglement production with finite-density kinetic reaction
activity.  It should, however, be understood as a channel-resolved
finite-density entanglement entropy, not as a thermodynamic entropy
of the bath or as the entanglement entropy of the full mixed many-body state.

As a proof-of-principle application, we evaluated this quantity in a
thermal quasiparticle bath of the $O(N)$ scalar extended model.  Within
the perturbative effective-potential and quasiparticle treatment used
for the benchmarks, the normalized ratio
$\widehat{\mathcal E}_{\rm fk}(T)$ is sensitive to the phase structure
obtained independently from the finite-temperature effective potential.
In the examples studied, $\widehat{\mathcal E}_{\rm fk}(T)$ develops a
finite discontinuity across first-order transitions, while in a
continuous-transition benchmark it remains continuous and its first
temperature derivative becomes nonanalytic.  This correlation has a
natural origin: the nonanalyticity of the thermal background is
transmitted to the branch-changing reaction-density kernel through dressed masses, phase-space and statistical weights, and background-dependent
amplitudes.  The scan results therefore
support the use of this collision-kernel observable as a phase-transition-type diagnostic.

The present analysis is limited to the leading $O(P)$ entanglement entropy,
$2\to2$ quasiparticle scattering, kinetic coarse-graining, and a
factorized background with diagonal momentum distributions.  Natural
extensions include $2\to n$ and higher-multiplicity processes,
higher-order coherence effects in the reduced flavor density matrix,
matrix-valued Wigner or kinetic density matrices for flavor and momentum
coherences, and more systematic treatments of thermal masses, widths,
screening effects, and non-equilibrium backgrounds.  These developments
would clarify the range of validity of the finite-density
scattering--entanglement dictionary and its possible use as a
quantitative diagnostic of phase structure and real-time kinetic
dynamics in early-universe and other thermal field-theory settings.

\vspace{12pt}
\begin{acknowledgments}

The work of J.L. is supported by the National Natural Science Foundation of China under Grants No.~12235001 and No.~12475103, and by the State Key Laboratory of Nuclear Physics and Technology under Grant No.~NPT2025ZX11. 
The work of X.P.W. is supported by the National Natural Science Foundation of China under Grant No.~12375095, and the Fundamental Research Funds for the Central Universities. The work of J.J.Z. is supported by the National Natural Science Foundation of China under Grants No.~11635001 and No.~11875072.  The authors gratefully acknowledge the valuable discussions and insights provided by the members of the Collaboration of Precision Testing and New Physics.
\end{acknowledgments}

\appendix
\setsectionprefixedsubsections

\section{Finite-density coherence blocks and reaction-density relation}
\label{app:finite-density-details}

The leading finite-density result used in Sec.~\ref{sec:finite-density}
is the diagonal, branch-changing part of a more complete reduced flavor
matrix.  This appendix records the minimal coherence-block derivation
behind Eq.~\eqref{eq:fd-theorem}.

For a fixed unordered incoming pair $\{a,b\}$, with the same
symmetry convention as in Eq.~\eqref{eq:reaction-density}, define
\begin{equation}
C_{cd;c'd'}
\equiv
\Tr_{\kin}
\left[
\ket{X_{cd}}
\bra{X_{c'd'}}
\right] .
\label{eq:fd-coherence-abstract}
\end{equation}
Here $\ket{X_{cd}}$ denotes the branch-projected outgoing
component generated by $ ab\to cd$ after averaging over the sampled
incoming momenta. It is a block label rather than an additional bath state.

Using Eq.~\eqref{eq:fd-channel-probability}, the diagonal entries are
\begin{equation}
C_{cd;cd}
=
P_{cd,ab}^{\rm FD}
=
S_{ab}\,\frac{\tau}{n_a n_b V}\,
\Gamma_{ab\to cd} .
\label{eq:fd-coherence-diagonal}
\end{equation}
Thus the same positive directed kernel that defines a diagonal
flavor--kinetic entropy component is the reaction density entering
kinetic equations.

Off-diagonal entries connect final branches with a common on-shell mass
multiset, their kinetic configurations identified through the common
phase-space measure of Eq.~\eqref{eq:common-phase-space}.  Their statistical
weight reduces to $\Xi_c\Xi_d$ on the diagonal, while for an
occupation-diagonal bath the spectator occupations record which final channel
occurred and suppress the off-diagonal entries. Their explicit form is not
needed in what follows.

The leading entropy is insensitive to these off-diagonal blocks.  By
positivity of the branch-space matrix,
\begin{equation}
\left|C_{cd;c'd'}\right|^2
\le
C_{cd;cd}
C_{c'd';c'd'} ,
\end{equation}
so off-diagonal coherences can enter the purity only through quadratic
combinations of transition probabilities.  Therefore, in the
finite-density $2\to2$ sector considered in the main text,
\begin{align}
\mathcal E_{\rm fk}^{\rm FD,(ab)}
&=
2
\sum_{cd\in\mathcal{BC}(ab)}
C_{cd;cd}
+
O(P^2)
\notag\\
&=
\frac{2S_{ab}\,\tau}{n_a n_b V}
\sum_{cd\in\mathcal{BC}(ab)}
\Gamma_{ab\to cd}
+
O(P^2) .
\label{eq:fd-coherence-leading-entropy}
\end{align}
Equation~\eqref{eq:fd-coherence-leading-entropy} is the
coherence-block form of Eq.~\eqref{eq:fd-theorem}: the finite-density
entanglement entropy keeps the positive branch-changing collision activity,
while the Boltzmann equation combines the same directed kernels into
net gain-minus-loss abundance flow.

\section{\texorpdfstring{$O(N)$}{O(N)} model details and numerical procedure}
\label{app:on-numerics}
\begingroup
\setlength{\abovedisplayskip}{3pt}
\setlength{\belowdisplayskip}{3pt}
\setlength{\abovedisplayshortskip}{2pt}
\setlength{\belowdisplayshortskip}{2pt}
\setlength{\jot}{2pt}

This appendix summarizes the thermal-potential input and the
finite-density scattering calculation used in
Sec.~\ref{sec:thermal-application}.  The effective potential is used to
locate the thermal background and to provide an independent reference
classification of the transition. The finite-density entanglement entropy
is then evaluated from the collision kernel of Sec.~\ref{sec:finite-density}.

At zero temperature,
\begin{align}
v_0
&=
\sqrt{\frac{6\mu_h^2}{\lambda_h}}
\simeq246~{\rm GeV},
\notag\\
m_h^2
&=
2\mu_h^2
=
\frac{\lambda_h}{3}v_0^2
\simeq(125~{\rm GeV})^2,
\qquad
\lambda_h=\frac{3m_h^2}{v_0^2}.
\label{eq:on-zero-temperature-input}
\end{align}

Along the background direction $\phi=\varphi$, $S_a=0$, the
field-dependent scalar masses are
\begin{align}
M_h^2(\varphi)
&=
-\mu_h^2+\frac{\lambda_h}{2}\varphi^2,
\notag\\
M_S^2(\varphi)
&=
\mu_S^2+\frac{\lambda_{hS}}{2}\varphi^2 .
\label{eq:on-field-dependent-masses}
\end{align}
The label $h$ denotes the radial field direction, while in the symmetric
phase the corresponding excitation is denoted by $\phi$ in the
channel labels of Eq.~\eqref{eq:on-symmetric-channels}.

The dressed one-loop potential follows the finite-temperature
effective-potential convention of Ref.~\cite{Curtin:2016urg},
restricted to the real-scalar $O(N)$ model.  Gauge-boson, fermion,
and complex-scalar terms absent from the model are not included:
\begin{align}
\bar M_i^2(\varphi,T)
&=
M_i^2(\varphi)+\Pi_i(T),
\notag\\
V_{\rm eff}^{\rm dressed}(\varphi,T)
&=
V_0(\varphi)
+
\sum_{i=h,S}
 g_i
\left[
V_{\rm CW}^i\!\left(\bar M_i^2\right)
\right.
\notag\\
&\qquad\qquad\qquad\left.
+
V_T^i\!\left(\bar M_i^2,T\right)
\right] .
\label{eq:on-dressed-potential}
\end{align}
Here $V_0(\varphi)$ is the tree-level potential of Eq.~\eqref{eq:on-potential} evaluated along $\phi=\varphi$, $\vec S=0$, and $\bar M_r^2(\varphi,T)$ are the dressed masses used in Eq.~\eqref{eq:on-bose-distribution}.
Here $g_h=1$ and $g_S=N$, corresponding to the same field content as the microscopic flavor sums in
Eqs.~\eqref{eq:on-active-symmetric} and~\eqref{eq:on-active-broken}.
For the scalar thermal self-energies we use
\begin{align}
\Pi_\phi(T)
&=
T^2
\left(
\frac{\lambda_h}{24}
+
\frac{N\lambda_{hS}}{24}
\right),
\notag\\
\Pi_S(T)
&=
T^2
\left(
\frac{\lambda_{hS}}{24}
+
\frac{(N+2)\lambda_S}{12}
\right) .
\label{eq:on-thermal-self-energies}
\end{align}
The self-energy $\Pi_\phi$ is used for the radial $h$ direction
and for the symmetric-phase $\phi$ excitation.

For each parameter point $(\mu_S,\lambda_{hS},\lambda_S)$, the
thermal background is obtained by minimizing
$V_{\rm eff}^{\rm dressed}(\varphi,T)$ in $\varphi$.  The global
minimum is denoted by $v(T)$.  A first-order transition is identified
by the existence of a critical temperature $T_c$ at which two
separated local minima are degenerate,
\begin{equation}
V_{\rm eff}^{\rm dressed}(0,T_c)
=
V_{\rm eff}^{\rm dressed}(v_c,T_c),
\qquad
v_c\neq0 .
\label{eq:on-fopt-condition}
\end{equation}
A continuous transition is identified when the thermal minimum connects
to the origin without a separated degenerate minimum.  This
potential-based classification is used only to locate the transition
and to provide the reference boundary overlaid in the scan plots. The
entanglement entropy is computed independently from the finite-density
reaction densities.

The occupation functions are Bose--Einstein distributions,
\begin{equation}
f_r(\mathbf p;T)
=
\frac{1}{\exp\!\left(\sqrt{\mathbf p^2+\bar M_r^2(v(T),T)}/T\right)-1}.
\label{eq:on-bose-distribution}
\end{equation}
Since all active species in this benchmark are scalars,
Eq.~\eqref{eq:stat-factor} gives $\Xi_r=1+f_r$, with no Pauli-blocking factor.

The finite-density amplitudes $\M_{cd,ab}^{\rm FD}$ are obtained from
the scalar interactions after expanding Eq.~\eqref{eq:on-potential}
around the thermal background.  In the symmetric phase the active
families in Eq.~\eqref{eq:on-symmetric-channels} are contact channels.
In the broken phase, $\phi=v(T)+h$, the shifted interactions include
the $hS_iS_i$ cubic vertex when exchange diagrams are retained.
Medium dependence enters through quasiparticle masses, statistical
factors, and $\M^{\rm FD}$ in Eq.~\eqref{eq:reaction-density}.
For each degenerate $O(N)$ family, the numerical rate is obtained from
one representative microscopic amplitude with the same vertex and symmetry
conventions, and the explicit sums in
Eqs.~\eqref{eq:on-active-symmetric} and~\eqref{eq:on-active-broken}
supply the $N$ or $N(N-1)$ multiplicity.

Numerically, $T_c$ and the phase label are first obtained from the
effective-potential scan.  At the temperatures entering
$\widehat{\mathcal E}_{\rm fk}(T)$ and $\mathcal J_{\rm fk}$, the
background, masses, distributions, amplitudes, and directed reaction
densities are evaluated consistently at the same $T$.  The
per-pair probabilities from Eq.~\eqref{eq:fd-channel-probability}
are then inserted into Eqs.~\eqref{eq:on-active-symmetric}
and~\eqref{eq:on-active-broken}. All incoming pairs carry $S_{ab}=2$. The $O(N)$ degeneracy allows one
representative channel per family to be multiplied by $N$ or
$N(N-1)$ 
The overall factor $\tau/V$ is common to all channel probabilities
and cancels in the normalized ratio $\widehat{\mathcal E}_{\rm fk}$
and in the signed jump $\mathcal J_{\rm fk}$.

\endgroup

\bibliographystyle{apsrev4-2}
\bibliography{references}

\end{document}